\documentclass[12pt]{article}

\usepackage{scicite}
\usepackage{times}

\newenvironment{sciabstract}{%
\begin{quote} \bf}
{\end{quote}}

\usepackage{amssymb,amsmath}
\usepackage{graphicx}
\usepackage{bm}				

\title{Orbital symmetries of charge density wave order in YBa\textsubscript{2}Cu\textsubscript{3}O\textsubscript{6+x}} 

\author{
Christopher McMahon,$^{1}$ A. J. Achkar,$^{1}$ E. H. da Silva Neto,$^{2,3,4,5,6}$\\
 I. Djianto,$^{1}$ J. Menard,$^{1}$ F. He,$^{7}$, R. Sutarto,$^{7}$ R. Comin,$^{8}$ \\
 Ruixing Liang,$^{3,9}$ D. A. Bonn,$^{2,3,9}$ W. N. Hardy,$^{3,9}$\\
 A. Damascelli,$^{2,3,9}$ D. G. Hawthorn$^{1,3,\ast}$\\
\\
\normalsize{$^{1}$Department of Physics and Astronomy, University of Waterloo, Waterloo, N2L 3G1, Canada}\\
\normalsize{$^{2}$Quantum Matter Institute, University of British Columbia, Vancouver,}\\
\normalsize{British Columbia V6T 1Z4, Canada}\\
\normalsize{$^{3}$CIFAR, Toronto, Ontario M5G 1Z8, Canada}\\
\normalsize{$^{4}$Max Planck Institute for Solid State Research, Heisenbergstrasse 1, D-70569 Stuttgart, Germany}\\
\normalsize{$^{5}$Department of Physics, University of California, Davis, California, 95616, USA}\\
\normalsize{$^{6}$Department of Physics, Yale University, New Haven, Connecticut 06511, USA}\\
\normalsize{$^{7}$Canadian Light Source, Saskatoon, Saskatchewan, S7N 2V3, Canada}\\
\normalsize{$^{8}$Department of Physics, Massachusetts Institute of Technology, Cambridge, MA, USA}\\
\normalsize{$^{9}$Department of Physics \& Astronomy, University of British Columbia, Vancouver,V6T 1Z1, Canada}\\
\\
\normalsize{$^\ast$To whom correspondence should be addressed; E-mail:  dhawthor@uwaterloo.ca.}
}

\date{}

\begin{document}

\baselineskip24pt

\maketitle 


\begin{sciabstract}

Charge density wave (CDW) order has been shown to compete and coexist with superconductivity in underdoped cuprates. Theoretical proposals for the CDW order include an unconventional $d$-symmetry form factor CDW, evidence for which has emerged from measurements, including resonant soft x-ray scattering (RSXS) in YBa$_2$Cu$_3$O$_{6+x}$ (YBCO). Here, we revisit RSXS measurements of the CDW symmetry in YBCO, using a variation in the measurement geometry to provide enhanced sensitivity to  orbital symmetry. We show that the $(0\ 0.31\ L)$ CDW peak measured at the Cu $L$ edge is dominated by an $s$ form factor rather than a $d$ form factor as was reported previously. In addition, by measuring both $(0.31\ 0\ L)$ and $(0\ 0.31\ L)$ peaks, we identify a pronounced difference in the orbital symmetry of the CDW order along the $a$ and $b$ axes, with the CDW along the $a$ axis exhibiting orbital order in addition to charge order.

\end{sciabstract}

\subsection*{INTRODUCTION}

While the presence of charge density wave (CDW) order in the
cuprates appears to be ubiquitous\cite{Tranquada1995,Wu2011,Ghiringhelli2012,Chang2012,Achkar2012,Blackburn2013,Comin2014,DaSilvaNeto2014,DaSilvaNeto2015,Tabis2014,Kohsaka2007}, open questions remain
about the microscopic character of the CDW order and whether it
is also generic. In particular, theoretical studies\cite{Metlitski2010,Sachdev2013,Efetov2013,Seo2007,Vojta2008,Atkinson2015,Lawler2010,Li2006,Allais2014,Chowdhury2014} have predicted
the CDW charge modulation to have a $d$ form factor in contrast
to a more conventional $s$ ($s^\prime$) form factor. Whereas a $s$ ($s^\prime$) CDW
involves a simple, monopolar sinusoidal modulation of the charge
density on the Cu (O) sites, a $d$ form factor CDW involves a quadrupolar
modulation on the bonds between Cu atoms, namely, the O
sites, that are out of phase for bonds oriented along $x$ and $y$, giving
a form factor with $d_{x^2-y^2}$ symmetry (see Fig. 1A). Such a $d$ form
factor CDW has been observed in Bi$_2$Sr$_2$CaCu$_2$O$_{8+x}$ (Bi2212) and Na$_x$Ca$_{2-x}$CuO$_2$Cl$_2$ (NCCOC) using scanning
tunneling microscopy (STM)\cite{Fujita2014}. Further evidence of a dominant
$d$ form factor CDW order was provided by nonresonant hard
x-ray scattering in YBa$_2$Cu$_3$O$_{6+x}$ (YBCO)\cite{Forgan2015} and by resonant
soft x-ray scattering (RSXS) measurements at the Cu L edge of the
$(0\ 0.31\ 1.48)$ CDW Bragg peak in YBCO\cite{Comin2015a}. However, the ability
of the RSXS measurements to distinguish between dominant $d$ and
dominant $s$ or $s^\prime$ factor CDW orders was close to the experimental
accuracy, providing some ambiguity to the conclusion of a dominant
$d$ form factor CDW order in YBCO. In contrast to these observations,
RSXS measurements at the Cu $L$ and O $K$ edges in the spin-charge
stripe ordered cuprate La$_{1.875}$Ba$_{0.125}$CuO$_4$ (LBCO) find the
CDW order to have predominantly $s$/$s^\prime$ form factor\cite{Achkar2016}, indicating
that the symmetry of the CDW form factor may not be generic
to the different cuprate materials. Moreover, Achkar et al.\cite{Achkar2016} also
showed that the orbital symmetry of CDW order is not generic even
within YBCO, differing for CDW order propagating along the $a$
and $b$ axes.

In this study, we present new RSXS measurements of the orbital
symmetry of CDW in YBCO. We follow an experimental approach
similar to \cite{Comin2015a} and \cite{Achkar2016} but with two key advances: (i) We use a
different experimental geometry that provides much greater contrast
to the $d$ versus $s^\prime$ or $s$ form factor than past work and (ii) use
this technique to study both the CDW order propagating along the
$a$ and $b$ axes by investigating the $(0.31\ 0\ L)$ and $(0\ 0.31\ L)$ Bragg
peaks, respectively. With the enhanced sensitivity to $d$ versus $s^\prime$ or $s$
form factors, we find no clear evidence for a $d$ form factor CDW in
YBCO, contrary to the conclusions of Comin et al. \cite{Comin2015a}. Rather,
measurements of the $(0\ 0.31\ L)$ peak at the Cu $L$ appear to be dominated
by an $s$ form factor component of the CDW order, similar to
LBCO \cite{Achkar2016}. In addition, by studying both the $(0.31\ 0\ L)$ and $(0\ 0.31\ L)$
Bragg peaks, we detail how the density wave order in YBCO exhibits
a profound difference in orbital symmetry in addition to the previously
established unidirectional character in the CDW Bragg peak
intensity and correlation length \cite{Blackburn2013,Blanco-Canosa2013,Comin2015,Chang2016} or nuclear magnetic resonance
(NMR) line broadening \cite{Wu2015}. Specifically, we show that the
density wave order involves orbital order in addition to the established
charge order but only for CDW order propagating along the
$a$ axis and not along the $b$ axis. Two orbital order scenarios are discussed
involving modulations of states beyond the expected $3d_{x^2-y^2}$
orbitals or orbital rotations that break $bc$ and $ab$ plane symmetries,
such as an oscillation of the orbitals about the $b$ axis.

The experimental technique involves measuring the resonant x-ray
scattering at the Cu $L$ absorption edge ($\hbar\omega$ = 931.4 eV) and the CDW
wave vector. At this photon energy, the scattering intensity is sensitive
to modulations in the Cu $3d$ or core $2p$ states and has proven to be an
effective probe of CDW order in the cuprates. By varying the photon
polarization relative the crystallographic axes, resonant x-ray scattering
can also provide insight into the orbital symmetry of the density
wave order. This can be understood by considering the CDW
scattering intensity, I, on resonance in terms of tensor quantities,
akin to the index of refraction in an anisotropic medium

\begin{equation}
I(\vec Q, \hbar\omega, \vec\epsilon , \vec\epsilon\,^\prime)\propto\left|\vec\epsilon\,^\prime\cdot \hat F(\vec Q, \hbar\omega)\cdot\vec\epsilon\,\right|^2,
\end{equation}
where $\vec\epsilon$ and $\vec\epsilon\,^\prime$ are the incident and scattered polarization vectors, respectively, and 
\begin{equation}
\hat F(\vec Q,\hbar\omega)=\sum_j \hat f_{j} (\hbar\omega) e^{i\vec Q \cdot \vec r_j}=
\begin{bmatrix}
F_{aa} & F_{ab} & F_{ac} \\
F_{ba} & F_{bb} & F_{bc} \\
F_{ca} & F_{cb} & F_{cc}
\end{bmatrix}
\end{equation}
is a tensoral expression of the x-ray scattering structure factor. $\hat f_j$ is
the atomic scattering form factor tensor for an atom at site $j$, and
the symmetry of $\hat f_j$ follows from the local point group symmetry of
that site \cite{Haverkort2010}. Accordingly, the symmetry of $\hat F(\vec Q,\hbar\omega)$ relates to (but is
distinct from) the point group symmetry of individual sites and
how that symmetry is modulated by the density wave order.

For instance, a simple $s$ or $s^\prime$ form factor CDW order, corresponding
to a sinusoidal modulation of charge density (and properties
proportional to charge density), would have a scattering tensor
that has the same symmetry as the average ($\vec Q$ = 0 ) electronic structure.
For the CuO$_2$ planes of YBCO, where holes in Cu d$_x^2-y^2$ orbitals
dominate the electronic structure that is probed at 931.4 eV, this is
approximately $D_4h$ symmetry, and $\hat F(\vec Q,\hbar\omega)$ would be a diagonal
tensor with $F_{aa}\cong F_{bb}\gg |F_{cc}|$. In contrast, CDW order with a dominant
$d$ form factor may feature a symmetry with negative $F_{aa}$/$F_{bb}$,
indicative of a density modulation on $x$- and $y$-oriented “bonds”
that are out of phase \cite{Achkar2016,Comin2015a}. This could be seen at the Cu $L$ edge
due to shifting in the energy of the Cu $2p_x$ and $2p_y$ core states
\cite{Efetov2013,Comin2015a,Achkar2016} or more directly in the occupation of the O $2p$ state by
probing the CDW peak at the O $K$ edge \cite{Achkar2016}. Other symmetries
to $\hat F(\vec Q,\hbar\omega)$ may be signatures of a density wave order that involves
orbital order or magnetic order.

\begin{figure}
\centering
\includegraphics[width=4in]{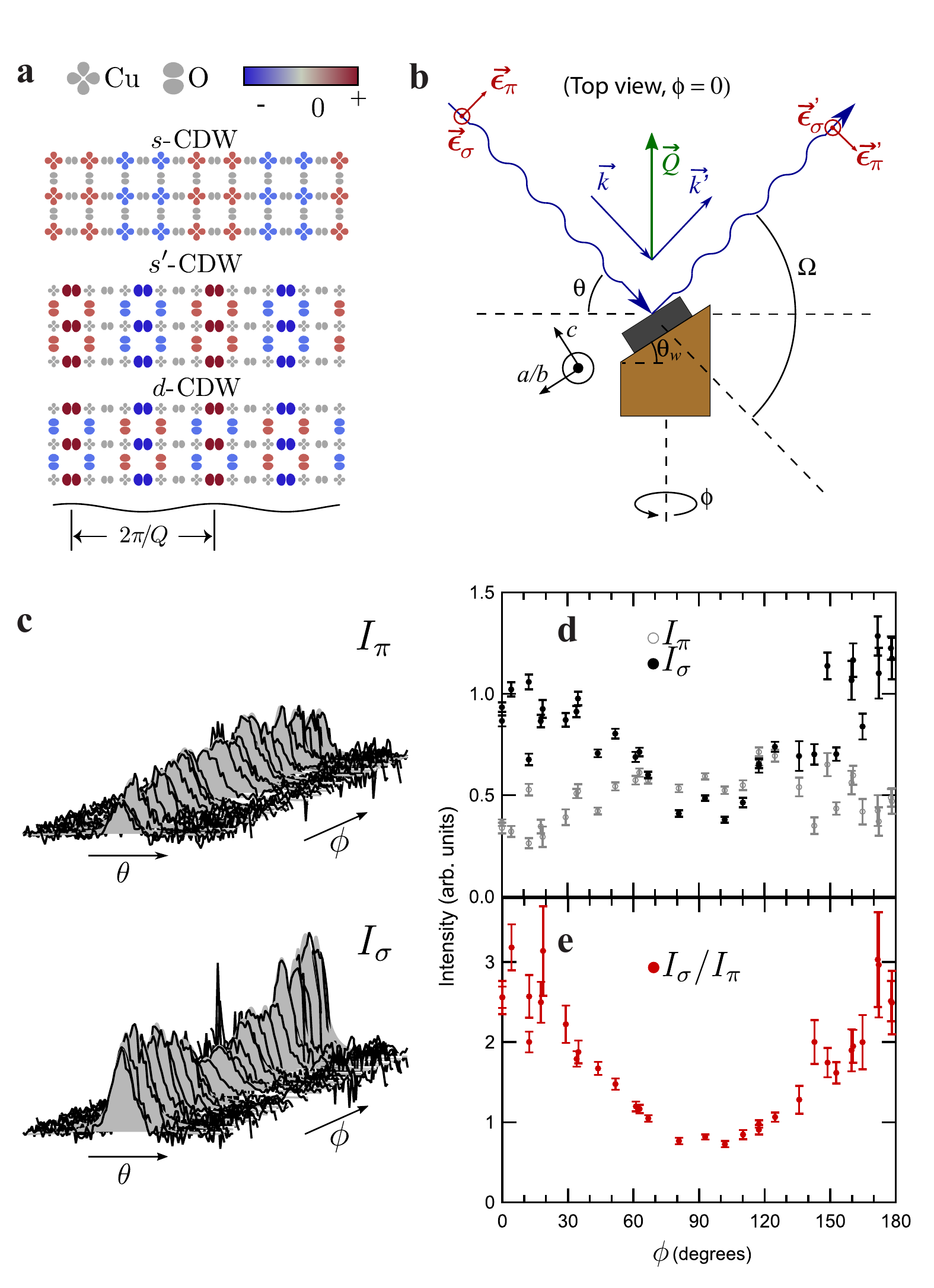}
\caption{Representations of CDW form factors and the experimental technique, resonant soft x-ray scattering, used to probe the CDW orbital symmetry. (a) Representation of the CuO$_2$ planes depicting $s$, $s^\prime$ and $d$ form factor CDW orders. (b) Schematic of the experimental geometry as seen from above showing the orientation when $\phi=0^\circ$.  As $\phi$ is rotated, the scattering vector $\vec Q$ remains unchanged, but the incident ($\vec \epsilon\,$) and scattered ($\vec \epsilon\,^\prime\,$) photon polarizations vary relative to the crystallographic axes.  (c) Intensity measurements of the CDW peaks for YBCO-6.75 at $(0\enspace 0.31\enspace 1.32)$. The peak profiles with fluorescent backgrounds removed as a function of $\theta$ and $\phi$ for $\pi$- and $\sigma$-polarizations.  Lorentzian fits to each peak are shown as grey shading.  The amplitudes $I_\sigma$ and $I_\pi$ (d) and ratio $I_\sigma/I_\pi$ (e) versus azimuthal angle.}
\label{Ipeaks}
\end{figure}

\begin{figure}
\centering
\includegraphics[width=\linewidth]{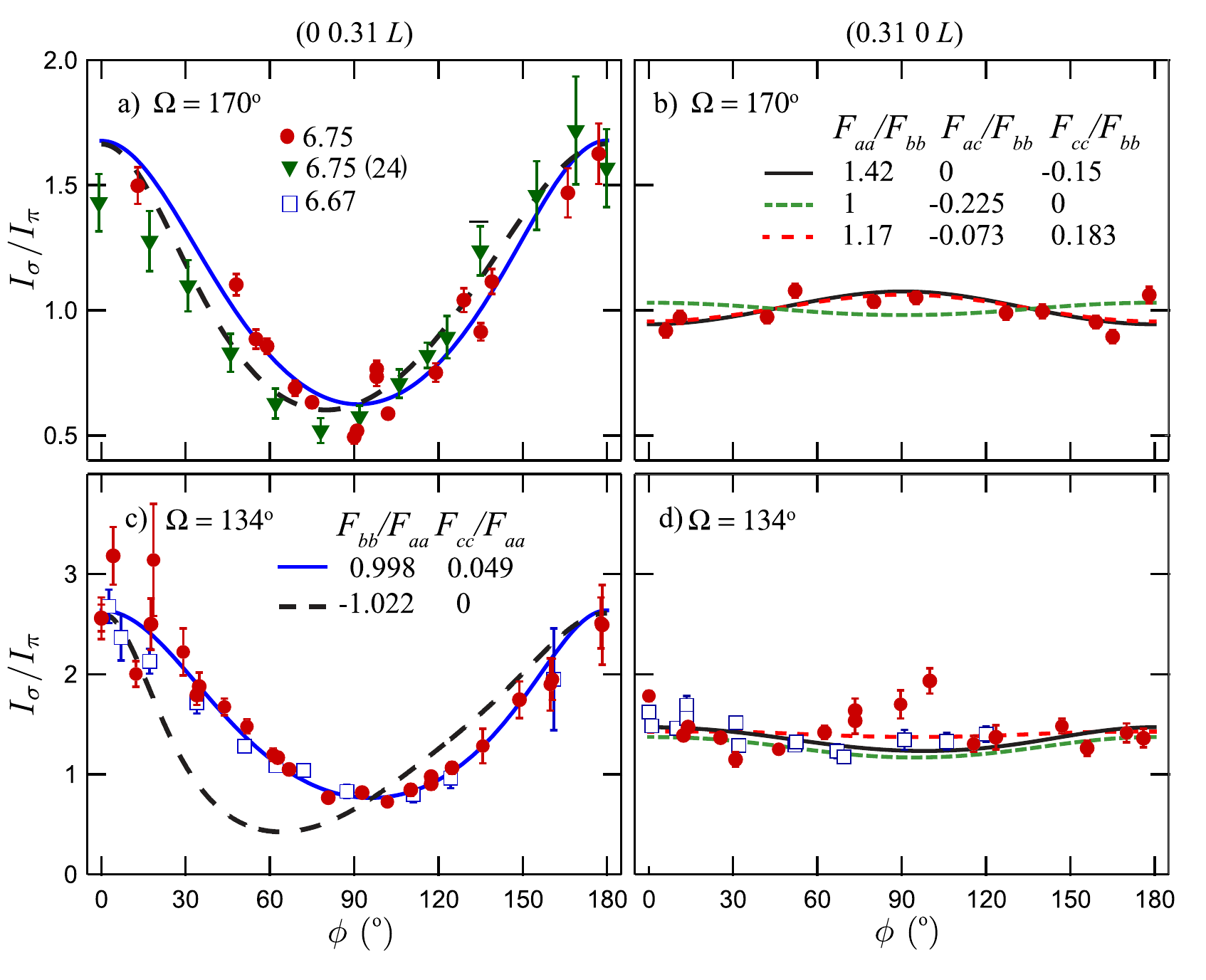}
\caption{The ratio of measured scattering intensities with $\sigma$ and $\pi$ incident polarization versus $\phi$ for the density wave order peaks at $(0\ 0.31\ L)$ and $(0.31\ 0\ L)$ for samples with O content 6.67 and 6.75.  $I_\sigma/I_\pi$ for $(0\ 0.31\ L)$ is shown in (a) and (c) with data from \cite{Comin2015a} reproduced in (a). $I_\sigma/I_\pi$ for $(0.31\ 0\ L)$ is shown in (b) and (d).  The top row (a and b) shows data taken at $\Omega\simeq170^\circ$ ($L=1.48$), while the bottom row (c and d) shows $\Omega\simeq134^\circ$ ($L=1.33$).  The lines are fits to model calculations.  For the ($0\ 0.31\ L)$ peak, a scattering tensor with $F_{aa}\simeq F_{bb}\gg F_{cc}$ agrees well with the data, whereas the dominant $d$ form factor model in \cite{Comin2015a} (dashed black) does not reproduce the $\Omega=134^\circ$ data.  For the $(0.31\ 0\ L)$ peak, a range of parameters corresponding to qualitatively different symmetries provide comparable quality fits.}
\label{FourPanel}
\end{figure}

Experimentally, the symmetry of $F\hat F(\vec Q,\hbar\omega)$ can be determined by
measuring the CDW Bragg peak intensity at a fixed $\vec Q_\text{CDW}$ and rotating
the sample azimuthally such that the orientation of $\vec\epsilon$ and $\vec\epsilon\,^\prime$ vary
relative to the crystallographic axes, as shown in Fig. 1 \cite{Comin2015a,Achkar2016}. This
approach has been used in analysis of the $(0\ 0.31\ 1.48)$ peak in
YBCO \cite{Comin2015a} to provide evidence that the CDW has a dominant $d$
form factor. However, the measurements by Comin et al. \cite{Comin2015a} use a
probing geometry with $\Omega\simeq170^\circ$ ($L\simeq1.5$), where the CDW peak
intensity is maximal. These measurements, while statistically favouring
a $d$ form factor, did not provide strong contrast between $s$, $d$,
and $s^\prime$ form factor models. Here, we revisit this analysis by measuring
the azimuthal dependence of the CDW order at $\Omega\simeq134^\circ$ ($L\simeq1.3$) in addition to $\Omega\simeq170^\circ$ ($L\simeq1.48$). Notably, the measurements
with $\Omega\simeq134^\circ$ ($L\simeq1.3$) probe the same CDW peak as $\Omega\simeq170^\circ$ ($L\simeq1.48$), owing to the fact that the quasi two-dimensional CDW peak
in YBCO is very broad in $L$. However, the variation in measurement
geometry affects how $\vec\epsilon$ and $\vec\epsilon\,^\prime$ span the crystallographic axes as $\phi$ is
rotated, yielding greater sensitivity to the sign of $F_{aa}$/$F_{bb}$, and thus
the form factor of the CDW order, for lower $\Omega$. In addition, we also
explore the orbital symmetry of the CDW order measured for both
the $(0.31\ 0\ L)$ and $(0\ 0.31\ L)$ peaks, representing CDW order propagating
along the $a$ and $b$ axes, respectively.

\subsection*{RESULTS}

The results of the measurements and subsequent fits for the $(0\ 0.31\ L)$
and $(0.31\ 0\ L)$ peaks using geometries with $L\simeq1.48$ and $L\simeq1.33$ are
shown in Fig. 2. In addition, data from Comin et al. \cite{Comin2015a} of the
$(0\ 0.31\ L)$ peak in YBCO-6.75 with $L\simeq1.48$ are reproduced in
Fig. 2A and are shown to be in good agreement with our measurements.
The first observation to make about the data in Fig. 2 is a
notable difference between measurements of the $(0.31\ 0\ L)$ and
$(0\ 0.31\ L)$ peaks, with the $(0\ 0.31\ L)$ peaks exhibiting substantially
larger dependence on $\phi$. We take this as compelling evidence to support
the existence of two different orbital symmetries for the CDW
order propagating along $a$ and $b$.

The data in Fig. 2 can be fit to determine the components of the
scattering and the underlying symmetry of the CDW order. Applying
this first to the $(0\ 0.31\ L)$ peak, we note that the fit parameters
from Comin et al. \cite{Comin2015a} used to argue a $d$ form factor CDW, shown
as the dashed black line in Fig. 2A, agree reasonably well with our
measurements for $\Omega=170^\circ$. However, this model does not agree
with our measurements at $\Omega=134^\circ$ (Fig. 2C), where a minimum in
$I_\sigma$/$I_\pi$ at $\phi\simeq60^\circ$ would be expected instead of the observed minimum at $\phi\simeq=100^\circ$. In contrast, a model with $F_{aa}\simeq F_{bb}\gg|F_{cc}|$ fits well to both $\Omega=170^\circ$ and $134^\circ$ measurements (a least squares fit gives $F_{bb}/F_{aa}=0.998\pm0.020$ and $F_{cc}/F_{aa}=0.049\pm0.041$). This symmetry for $\hat F(\vec Q,\hbar\omega)$
is consistent with the Cu $L$ edge measurements of the CDW order
being dominated by an $s$ form factor component of the CDW order,
which would entail a modulation in the orbital occupation (or
property proportional to the orbital occupation) of Cu $3d_{x^2-y^2}$
states. That is, for the $(0\ 0.31\ L)$ peak, the symmetry of the Cu orbitals
that are spatially modulated in the CDW is the same as the average
symmetry of the in-plane Cu. This symmetry ($F_{aa}\simeq F_{bb}\gg|F_{cc}|$) is
similar to that found in Cu $L$ edge measurements of the CDW order
in LBCO, where a predominant $s^\prime$ symmetry to the CDW order
was also deduced from measurements at the O $K$ edge \cite{Achkar2016}.

This result introduces the possibility that, similar to LBCO, the
CDW order in YBCO also has predominantly $s$/$s^\prime$ orbital symmetry,
differing from the $d$ form factor CDW order observed in Bi2212
and NCCOC by STM \cite{Fujita2014}. However, we note that the absence of
evidence for a $d$ form factor CDW order in the present measurements
does not fully rule out its relevance to YBCO. A $d$ form factor
density wave order is suggested by the pattern of oxygen displacements
deduced from nonresonant x-ray scattering measurements
\cite{Forgan2015}. Reconciling that result with our present observations may
require understanding details of how different experimental techniques
couple to form factor of the density wave order. For instance,
STM measurements in Bi2212 indicate that a $d$ form factor coexists
with $s$ and $s^\prime$ form factors and that the degree of $s$, $s^\prime$, and $d$ contributions
to the form factor depends on the sample bias, with a $d$ form
factor dominant at the pseudo-gap energy scale, but a substantial $s^\prime$
contribution at lower and higher bias \cite{Hamidian2016,Choubey2017}. That is, the $s$, $s^\prime$, and
$d$ contributions to the form factors depend on the energy and
momentum of electronic states. Similarly, it is anticipated that the
coupling of resonant x-ray scattering measurements to the symmetry
of the density wave order have dependence on the incident photon
energy \cite{Achkar2016}. While the $I_\sigma/I_\pi$ ratio in resonant x-ray measurements at $\phi=0$ does not appear to exhibit an energy dependence at the Cu $L$
edge \cite{Achkar2012}, indicating the symmetry probed is approximately energy
independent at the Cu $L$ edge, the electronic states and subsequent
sensitivity to the CDW form factor can vary between the Cu $L$ and
O $K$ edges or within the O $K$ edge \cite{Achkar2016}. Moreover, we anticipate that
measurements at the Cu $L$ edge are more sensitive to $s$ than $d$ or $s^\prime$
contributions to the form factor CDW \cite{Comin2015a,Achkar2016}. Future theoretical
work on the microscopic character CDW order may enable these
measurements place quantitative constraints on the degree of $d$ form
factor density wave order in YBCO.

We now turn our attention to fitting the symmetry of $\hat F(\vec Q,\hbar\omega)$
for the $(0.31\ 0\ L)$ peak. In contrast to the $(0\ 0.31\ L)$ peak, fits of the
symmetry of $\hat F(\vec Q,\hbar\omega)$ show a substantial departure from the average
point group symmetry of the CuO$_2$ planes for the $(0.31\ 0\ L)$ peak.
This indicates significant orbital ordering in addition to charge order
for the density wave orders propagating along the $a$ axis that is not
observed (i.e., not present or too small to distinguish) along the $b$
axis. Unfortunately, the present measurements are unable to uniquely
determine the symmetry of $\hat F(\vec Q,\hbar\omega)$. Rather, as shown for selected
fits in Fig. 2 (B and D), a range of parameters/symmetries with different
physical interpretations provide adequate fits to the data (the
regions of fit parameters with comparable reduced $\chi^2$ are shown in
the Supplementary Materials). This analysis reveals two symmetry
distinct scenarios that are found to fit well to the data: models having
substantial in-plane orbital asymmetry $F_{aa} > F_{bb}$ and/or models having
off-diagonal $F_{ac} = F_{ca}\neq 0$ elements.

The first scenario, involving a substantial in-plane asymmetry,
$F_{aa} > F_{bb}$, may result from a modulation of the in-plane Cu electronic
structure that involves Cu $3d$ orbitals beyond simply the $3d_{x^2-y^2}$
states that are known to dominate the low-energy electronic structure
of the CuO$_2$ planes. Alternately, $F_{aa} > F_{bb}$ may be indicative of
a chain layer contribution to the scattering. However, analysis of the
energy dependence of the CDW resonant x-ray scattering did not
reveal a contribution to the CDW peak from Cu in the chain layer,
which resonates at different photon energies from the Cu in the
CuO$_2$ planes \cite{Achkar2012,Hawthorn2011b}. Moreover, the azimuthal dependence of the
$(0.31\ 0\ L)$ peak does not vary as a function of temperature (Fig. 3) or
doping (Fig. 2), ruling out contributions from ortho chain ordering
to the $(0.31\ 0\ L)$ peak.

In the second scenario, the sizeable off-diagonal $F_{ac} = F_{ca}$ terms
indicate a density wave order with substantial breaking of $bc$ and $ab$
plane mirror symmetries. This could occur if the orientation of the
unoccupied Cu $d$ orbitals oscillates about the $b$ axis such that the $bc$
and $ab$ plane mirror symmetry of an individual Cu site is broken and
modulated with period of the CDW. Such a state may be consistent
with the previously reported pattern of lattice displacements refined
from nonresonant hard x-ray scattering \cite{Forgan2015}. As depicted in Fig. 4,
these lattice displacements may result in both a modulation of the
orbital occupation and orbital orientation, which would be consistent
with the presence of finite $F_{ac,ca}$ terms in the scattering tensor.
However, since similar patterns of lattice displacements for CDW
order along the $a$ and $b$ axes are deduced from nonresonant x-ray
scattering \cite{Forgan2015}, we may have expected to observe off-diagonal terms
in the $(0\ 0.31\ L)$ as well.

The key question that arises from these results is why the
$(0.31\ 0\ L)$ and $(0\ 0.31\ L)$ peaks exhibit different orbital symmetries.
In many respects, the density wave orders along $a$ and $b$ appear similar,
having comparable correlation lengths, temperature dependencies,
energy dependence, intensities at 1/8 doping, and patterns
of lattice displacements \cite{Achkar2012,Blackburn2013,Forgan2015,Blanco-Canosa2013}. However, they also have key
differences, such as different doping dependence to their intensities
\cite{Blackburn2013,Blanco-Canosa2013} and response to applied magnetic field \cite{Chang2016}. Moreover,
NMR has shown differences in line broadening for O(2) and O(3)
sites (2), as well as differences between Cu sitting below full and
empty chains (29), that may relate the to the asymmetry observed
here. An explanation of these differences may lie in identifying the
origin of the different orbital symmetries of the density wave orders
along $a$ and $b$.

\begin{figure}
\centering
\includegraphics[width=0.5\linewidth]{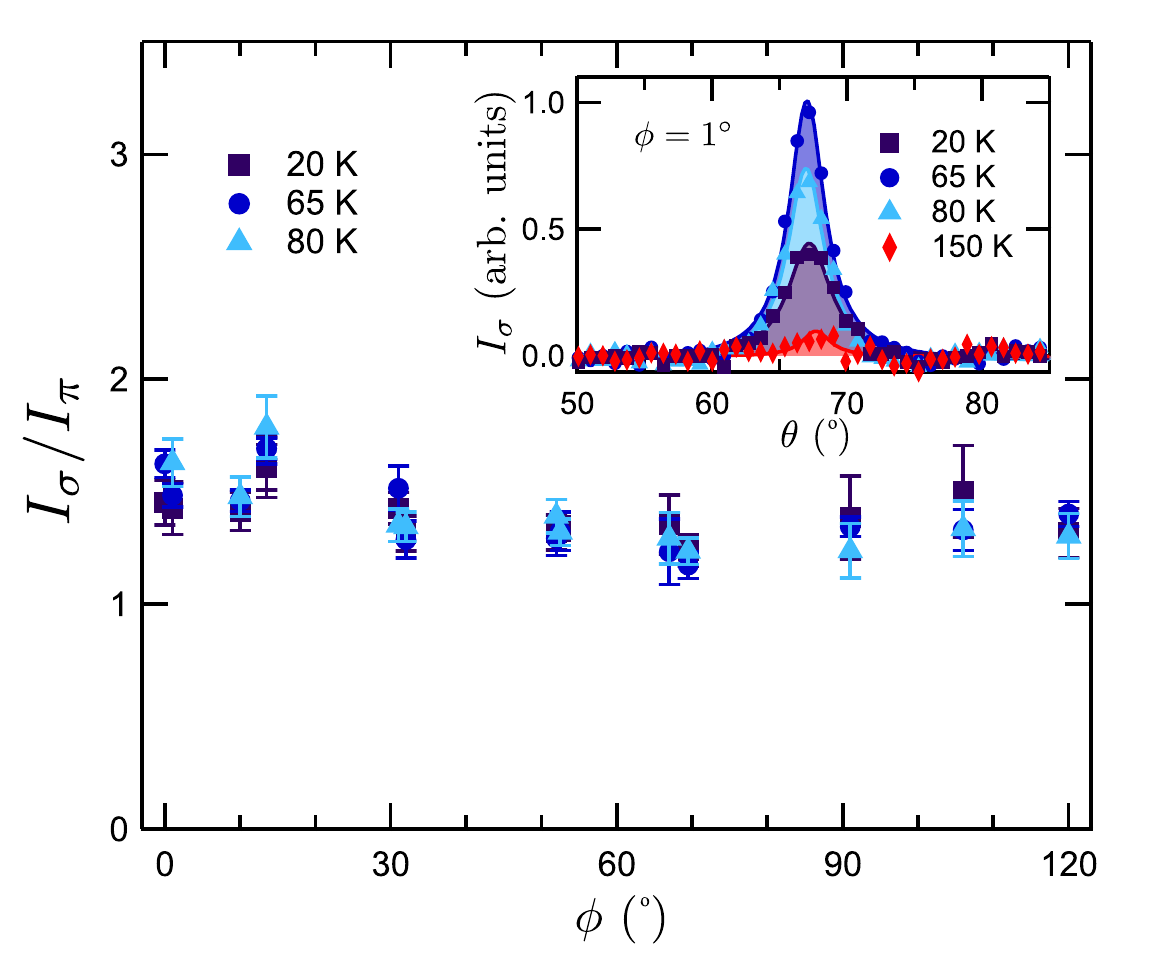}
\caption{Measurement of the  $(0.31\ 0\ 1.32)$ peak in YBCO-6.67 for a series of temperatures and azimuthal rotations, $\phi$.  $I_\sigma/I_\pi$ versus $\phi$ at different temperatures overlap within uncertainty and show no sign of a temperature dependence in the form factor. Inset: Scans of the intensity of the $(0.31\ 0\ 1.33)$ peak with $\sigma$ polarization at $\phi=1^\circ$ at various temperatures.}
\label{tempPlot}
\end{figure}

\begin{figure}
\centering
\includegraphics[width=0.5\linewidth]{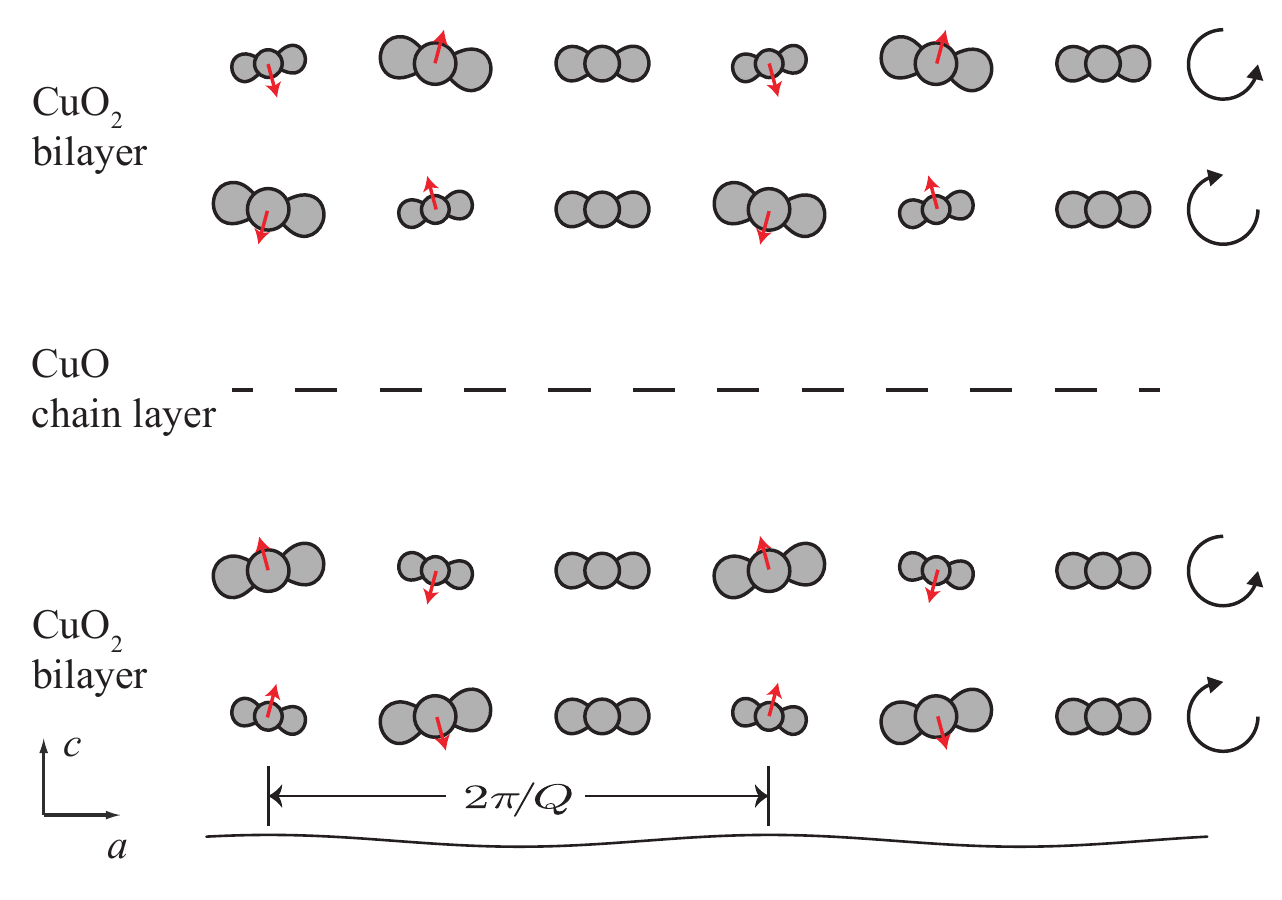}
\caption{A simplified pattern of modulated charge (depicted by the size of the Cu 3$d$ orbitals) and orbital symmetry that would give rise to off-diagonal, $F_{ac}$, elements in $\hat F(\vec Q,\hbar\omega)$.  The pattern of lattice displacements of the Cu atoms deduced from nonresonant x-ray diffraction is shown as red arrows\cite{Forgan2015}. These displacements trace out ellipses in the $ac$ plane as position advances along the $a$ axis.  These displacements (as well as the displacements of the neighbouring O, Y and Ba atoms) break and modulate the local $bc$ and $ab$ plane mirror symmetries.  This modulation can lead to an alternation in the orbital occupation (charge) of the Cu (roughly associated with the proximity of the in-plane Cu to the chain layer), and the direction of atomic displacements can lead to an alternation in the orientation of the unoccupied orbitals.  A more realistic model must incorporate the impact of O displacements on Cu 3$d$ orbital occupation.}
\label{modulation}
\end{figure}

\subsection*{MATERIALS AND METHODS}

Measurements were performed at the Resonant Elastic and Inelastic X-ray Scattering (REIXS) beamline of the Canadian
Light Source synchrotron \cite{Hawthorn2011a}. Samples were mounted on a copper
plug angled at $\theta_w=32.5^\circ$ and $35.5^\circ$ to achieve $L = 1.48$ and $L = 1.33$,
respectively, as depicted in Fig. 1A. The photon energy was set to
931.3 eV, the peak of the Cu $L_3$ absorption edge where the CDW
ordering peak is maximal \cite{Ghiringhelli2012,Achkar2012}. The incident photon polarization
was set to either $\sigma$ or $\pi$. However, the scattered photon polarization
or energy is not resolved. $H$ and $K$ scans at various $\phi$ values
were performed by rocking $\theta$ at fixed $\Omega$ such that the peak is
centered at a fixed $Q_\text{CDW}$ that does not vary with $\phi$. $\phi = 0$ is defined
in Fig. 1 and relates to $\alpha$ from \cite{Comin2015a} by $\alpha = 180^\circ-\phi$.

Two single-crystal samples of YBa$_2$Cu$_3$O$_{6+x}$ with oxygen stoichiometry
of $x = 0.75$ and 0.67, respectively, were measured using
the same samples from previous experiments in \cite{Achkar2016,Achkar2014}. The CDW
peak amplitudes were determined by first subtracting a fluorescent
background using a fifth-order polynomial and then fitting the CDW
peaks to a Lorentzian function, similar to \cite{Achkar2016}. Variations in the details
of the fitting procedure were explored and showed little effect
on the $\phi$ dependence of $I_\sigma/I_\pi$. Analysis of the symmetry of $\hat F(\vec Q,\hbar\omega)$
included the impact of the polarization dependence of the absorption
coefficient on the scattering intensity, as detailed in \cite{Achkar2016}. Figure 1
(A and B) shows the $(0\ 0.31\ L)$ peak intensities versus $\phi$ of the
YBCO-6.75 sample on the $35.5^\circ$ plug with backgrounds removed.




\bibliography{CDWSymmetryManuscript}
\bibliographystyle{Science_mod}

\noindent\textbf{Acknowledgments:}
We acknowledge discussions with S. Hayden, J. van Wezel, and G. A. Sawatzky.
\textbf{Funding:}
This work was supported by the Canada Foundation for Innovation
(CFI), CIFAR, and the Natural Sciences and Engineering Research Council of Canada
(NSERC). Part of the research described in this paper was performed at the Canadian Light
Source, a national research facility of the University of Saskatchewan, which is supported
by the CFI, the NSERC, the National Research Council (NRC), the Canadian Institutes of
Health Research (CIHR), the Government of Saskatchewan, and the University of
Saskatchewan. The work at UBC was supported by the Max Planck-UBC-UTokyo Centre for
Quantum Materials and the Canada First Research Excellence Fund, Quantum Materials
and Future Technologies Program. This research was undertaken thanks, in part, to
funding from the Killam, Alfred P. Sloan, and NSERC’s Steacie Memorial Fellowships (to
A.D.), the Alexander von Humboldt Fellowship (to A.D.), the Canada Research Chairs
Program (to A.D.), and CIFAR Quantum Materials Program.
\textbf{Author contributions:}
D.G.H.
and A.J.A. conceived of the experiments. C.M., A.J.A., E.H.d.S.N., I.D., F.H., R.S., and D.G.H.
performed the RSXS measurements. R.L., D.A.B., and W.N.H. provided the YBCO crystals.
D.G.H., C.M., A.J.A., J.M., and I.D. were responsible for data analysis. R.C., E.H.d.S.N., A.D.,
R.S., A.J.A., and C.M. discussed and developed the interpretation of the data and
contributed to the manuscript. C.M. and D.G.H. wrote the manuscript. D.G.H. is
responsible for overall project direction, planning, and management.
\textbf{Competing
interests:}
The authors declare that they have no competing interests.
\textbf{Data and
materials availability:}
All data needed to evaluate the conclusions in the paper are
present in the paper and/or the Supplementary Materials. Additional data related to this
paper may be requested from the authors.

\end{document}



\baselineskip24pt

\title{Supplementary information for:\\Orbital symmetries of charge density wave order in YBa$_2$Cu$_3$O$_{6+x}$} 

\date{}
\maketitle

\begin{figure}[h]
\centering
\resizebox{6.2in}{!}{\includegraphics{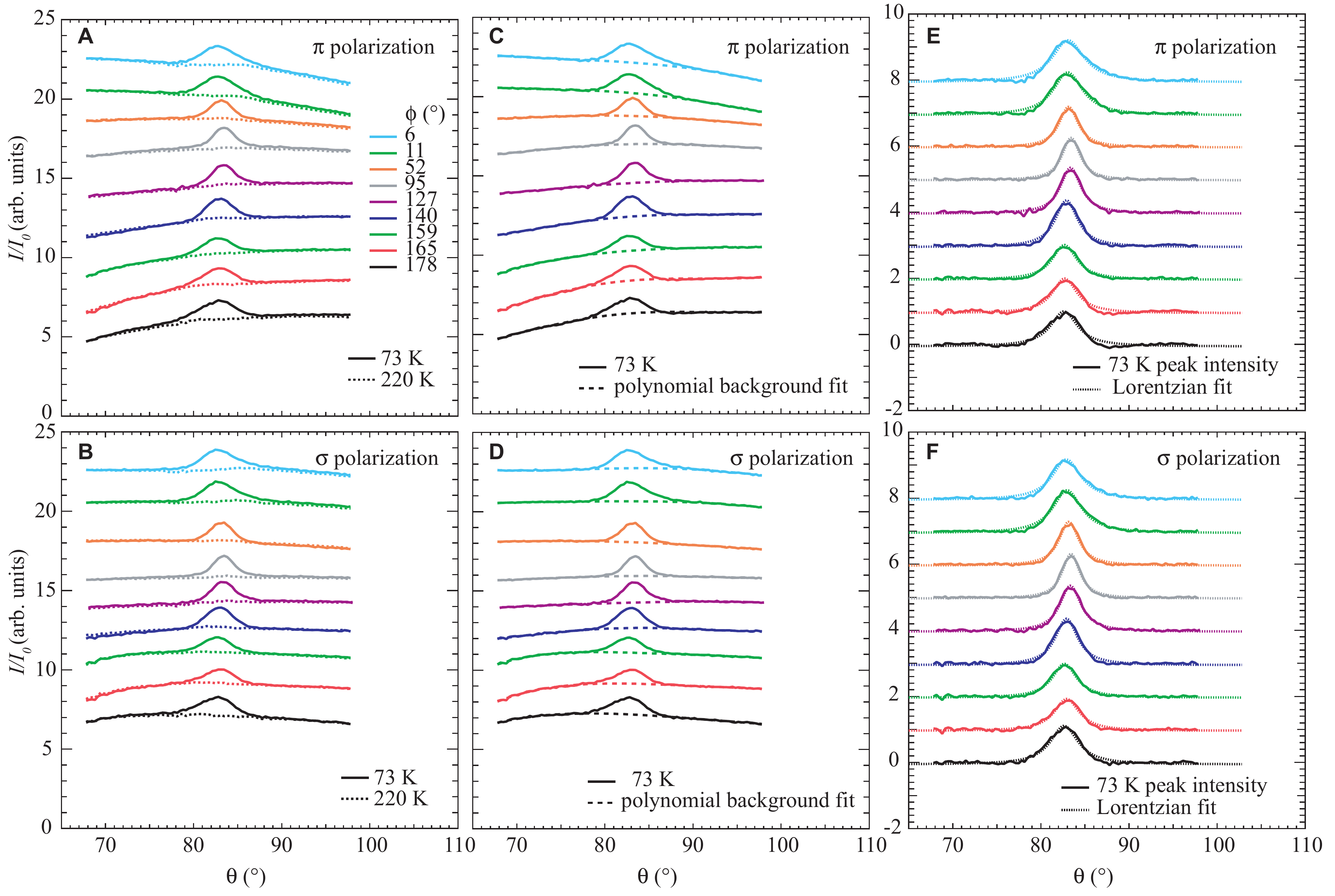}}
\caption{{\bf Example of the measurement and extraction of peak intensities.}  $\theta$ scans at various $\phi$ values through the CDW peak in the YBa$_2$Cu$_3$O$_{6.75}$ sample at (0.31 0 1.48) for A) $\pi$ and B) $\sigma$ incident photon polarization.  For clarity, data at different $\phi$ values are offset in the y-axis increments of 2 (Measurements at  $\phi = 178^\circ$ are not offset.  C) and D) fits to the fluorescent background of the data at 73 K using a 5th order polynomial.  E) and F) The peak intensity for various  $\phi$ values after subtraction of a polynomial background along with Lorentzian fits to the peaks.}
\label{figdata}
\end{figure}

In this Supplementary Information, we provide additional details regarding i) extracting the ratio of scattering intensity for $\sigma$ and $\pi$ incident photon polarization from the experimental data and ii) assess the range of fit parameters that provide good agreement with the measurements by examining the reduced-$\chi^2$ of model calculations.  

\section*{Experimental data.}

In fig.~\ref{figdata} we provide an example of the experimental data that was used to determine the $\phi$ dependences of $I_\sigma$ and $I_\pi$ shown in the main text.  The peak intensities are determined by scanning $\theta$ through the CDW peak position.  As shown in fig.~\ref{figdata} A and B, at 220 K, above the CDW ordering temperature, a smooth fluorescent background is observed.  At 73 K a clear CDW peak is observed in addition to the fluorescent.  Measurements both above and below the CDW ordering temperature were not performed at for all samples and peaks.  As such, in order to subtract the fluorescent background and determine the peak intensity, we fit a 5th order polynomial to the background (fig.~\ref{figdata} C and D), which is subtracted to reveal the CDW scattering intensity (fig.~\ref{figdata} E and F).  The resulting CDW peaks are fit to a Lorentzian line shape to determine the peak intensity vs. $\phi$ for $\sigma$ and $\pi$ incident photon polarization and determine their ratio $I_\sigma/I_\pi$, as shown in figure~1 and 2 of the main text.

\section*{Assessment of the quality of fits and range of parameters that agree with the measurements.}
For the (0 0.31 0 $L$) peak, the measured ratio $I_\sigma/I_\pi$ was fit to a model with; 
\begin{equation}
\hat F(\vec Q,\hbar\omega)\sim\
\begin{bmatrix}
F_{aa} & 0 & 0 \\
0& F_{bb} & 0 \\
0& 0 & F_{cc}
\end{bmatrix}
\end{equation}

\begin{figure}[h]
\centering
\resizebox{\columnwidth}{!}{\includegraphics{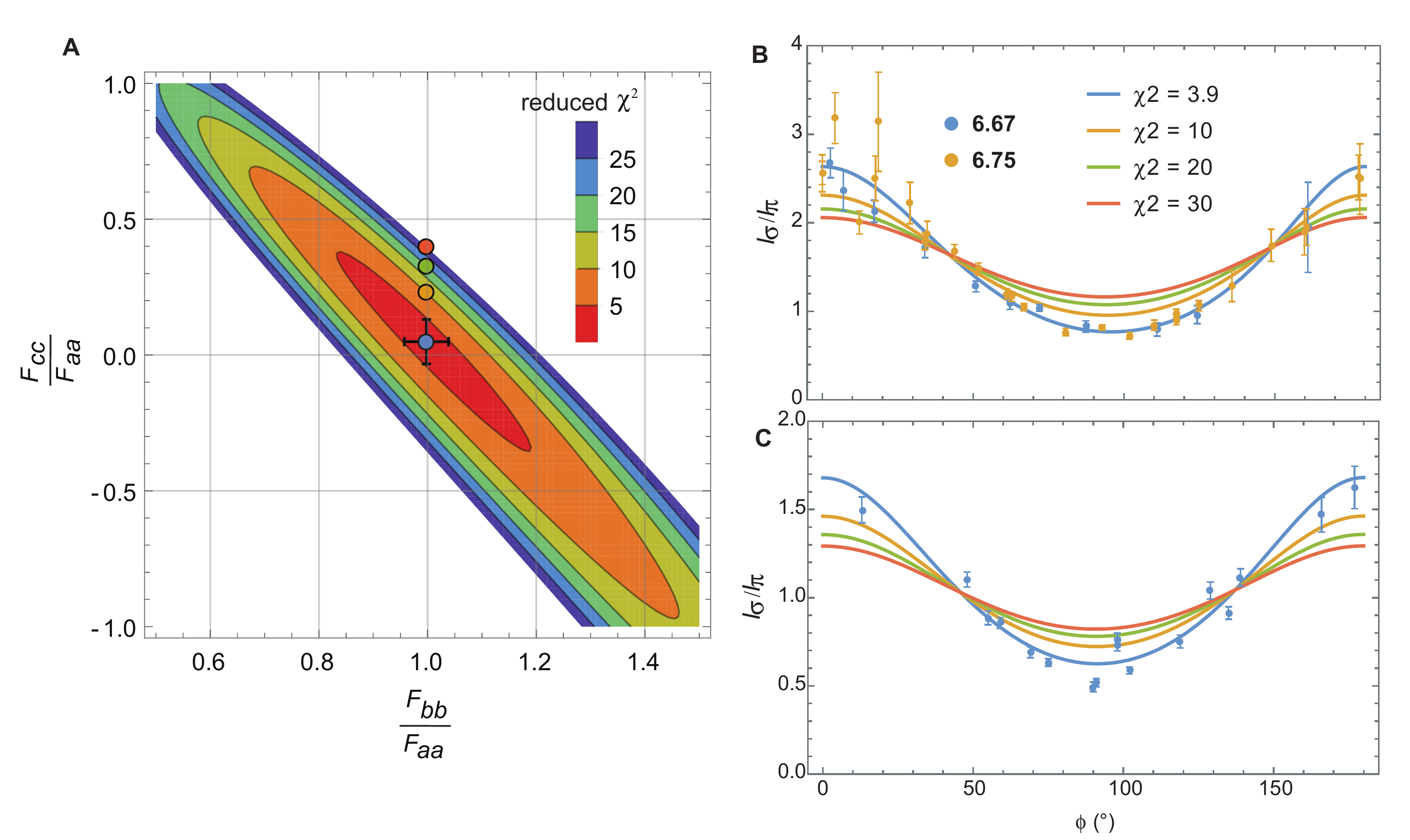}}
\caption{{\bf Assessment of the fit quality for the (0 0.31 $L$) peak.}  {\bf A} A map of the reduced $\chi^2$ from fits to data at both (0.31 0 1.48) and (0.31 0 1.32) vs.$F_{cc}/F_{aa}$, and $F_{bb}/F_{aa}$. The blue circle represents the best fit value.  {\bf B} and {\bf C} Examples of $I_\sigma/I_\pi$ vs. $\phi$ calculated for different reduced $\chi^2$ values. The curves in {\bf B} and {\bf C} were calculated using  parameters given by the point in {\bf A} of the same colour as the curve.}
\label{Kpeakchi2}
\end{figure}

The best fit to the data was achieved with $F_{bb}/F_{aa} = 0.998 \pm 0.020$ and $F_{cc}/F_{aa} = 0.049 \pm 0.041$.  The standard deviation, however, does not adequately represent the uncertainty in the parameters. This is because $F_{bb}/F_{aa}$ and $F_{cc}/F_{aa}$ are not completely independent fitting parameters.  Rather, variation in one parameter can be offset with variation in the other in order to improve the fit.  A better assessment of the range of parameters that provide good agreement with the data can be achieved by examining the dependence on fit parameters of the reduced $\chi^2$ statistic, as shown in Figure~\ref{Kpeakchi2} A). Good fits to the data are found in an elliptical range of parameters around the best fit value, where $F_{aa} \simeq F_{bb} << F_{cc}$ with a level of agreement comparable to the scatter in the data are roughly found for range of parameters inside the $\chi^2_0 = 5$ contour.  A comparison between the measurement and model calculations with selected parameters (shown by the solid circles in fig.~\ref{Kpeakchi2} A) that have different values of $\chi^2_0$ is shown in Fig.~  ~\ref{Kpeakchi2} B and C.  

For the (0.31 0  0 $L$) peak, the measured ratio $I_\sigma/I\pi$ was fit to a more a general model with off-diagonal terms $F_{ac} = F_{ca} $, indicative of broken $bc$ and $ab$ plane mirror symmetries:
\begin{equation}
\hat F(\vec Q,\hbar\omega)\sim\
\begin{bmatrix}
F_{aa} & 0 & F_{ac}\\
0& F_{bb} & 0 \\
F_{ac}& 0 & F_{cc}
\end{bmatrix}
\end{equation}

\begin{figure}[h]
\centering
\resizebox{5in}{!}{\includegraphics{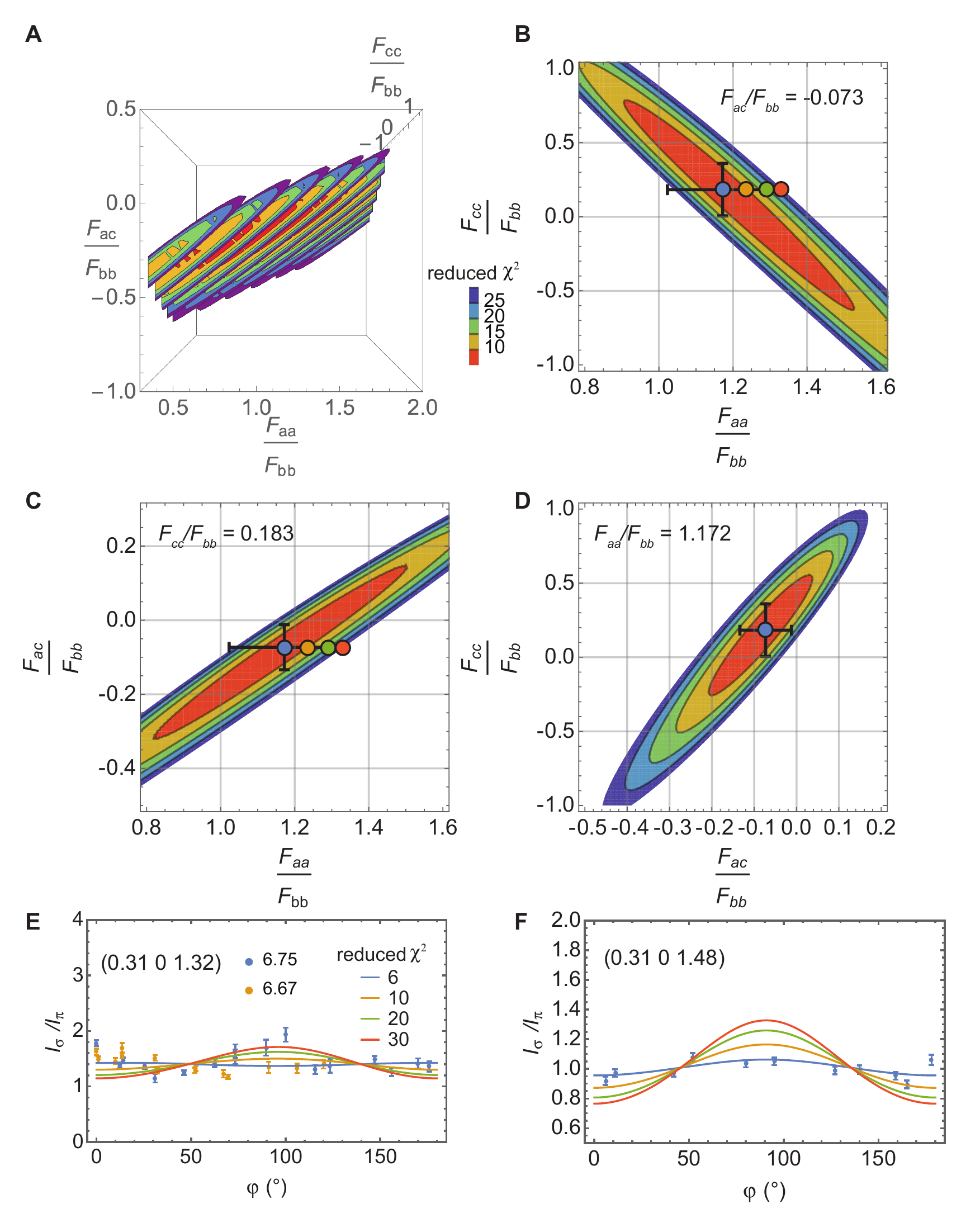}}
\caption{{\bf Assessment of the fit quality for the (0.31 0 $L$) peak.}  {\bf A} A map of the reduced $\chi^2$ from fits to the 6.75 data at both (0.31 0 1.48) and (0.31 0 1.32) vs. $F_{ac}/F_{bb}$, $F_{cc}/F_{bb}$, and $F_{aa}/F_{bb}$.  Contours of constant $\chi^2$ are shown are shown as slices through the 3D parameter space. {\bf B}, {\bf C} and {\bf D} 2D slices of reduced-$\chi^2$ through the best fit value $F_{ac}/F_{bb}$ = -0.073, $F_{cc}/F_{bb}$ = 0.183, and $F_{aa}/F_{bb}$ = 1.172.  The blue circle represents the best fit value and the error bars denote the 95\% confidence interval of the fit.  {\bf E} and {\bf F} Examples of $I_\sigma/I_\pi$ vs. $\phi$ calculated for different reduced $\chi^2$ values.  The curves in {\bf E} and {\bf F} were calculated using  parameters given by the point in {\bf C} and {\bf D} of the same colour as the curve.
}
\label{Hpeakchi2case6}
\end{figure}

The best fit to the data on the 6.75 sample gave $F_{ac}/F_{bb}$ = -0.073, $F_{cc}/F_{bb}$ = 0.183, and $F_{aa}/F_{bb}$ = 1.172. In figure~\ref{Hpeakchi2case6} we present the variation of reduced $\chi^2$ statistic with model parameters.  As shown, a range of model parameters provide good agreement with the data with fits with reduced $\chi^2 < 10$ all provide similar quality, comparable to the scattering in the data. This region with $\chi^2 < 10$ includes models with $F_{ac} = 0$ and $F_{aa} /F_{bb}$ significantly greater than 1, models that retain $ab$ and $bc$ plane mirror symmetries but has significant in-plane asymmetry.  However, it also includes a model with $F_{aa} \simeq F_{bb}$ and $F_{ac} = -0.22$, a model that breaks $ab$ and $bc$ plane mirror symmetries but retains approximate in-plane asymmetry of the diagonal elements, similar to the (0 0.31 $L$) peak.

